\documentclass[12pt, a4paper]{article}
\usepackage[height=22cm,width=16cm, centering]{geometry}
\usepackage{setspace}

\usepackage[utf8]{inputenc}

\usepackage{amsmath,amssymb,mathtools}
\usepackage{graphicx}
\usepackage{subfigure}
\usepackage{braket}
\usepackage{color,comment}
\usepackage[sort&compress, numbers, merge]{natbib}

\definecolor{phthaloblue}{rgb}{0.0, 0.06, 0.54}
\definecolor{purple}{rgb}{0.5 ,0, 0.7}
\definecolor{bluegreen}{rgb}{0, 0.45, 0.35}
\definecolor{sakura}{rgb}{1 ,0.52, 0.74}
\definecolor{wakakusa}{rgb}{0.45 ,0.74, 0}
\definecolor{brown}{rgb}{0.48 ,0.23, 0}
\definecolor{skyblue}{rgb}{0.21 ,0.7, 1.}
\definecolor{purplegray}{rgb}{0.35,0.35,0.73}
\usepackage{hyperref}
\hypersetup{colorlinks=true, linkcolor=bluegreen, citecolor=purplegray, urlcolor=purplegray}

\allowdisplaybreaks

\begin{document}

\begin{titlepage}
\begin{center}
\leavevmode \\
\vglue -.5in

\noindent

\vglue .3in
  {\LARGE \bf Spacetime Instability and 
  Quantum Gravity \\ \vskip 3 mm as Low  Energy Effective Field Theory  }

\vglue .40in

{\large
Hiroki Matsui$^{a,}$\footnote{\tt
hiroki.matsui@yukawa.kyoto-u.ac.jp},
}

\vglue.3in

\textit{
${}^{a}$ \,  Center for Gravitational Physics and Quantum Information, Yukawa Institute for Theoretical Physics, Kyoto University, 606-8502, Kyoto, Japan
}

\end{center}

\vglue 0.25in

\begin{abstract}\normalsize
We discuss spacetime instability 
for effective field theories of quantum gravity.
The effective action of gravity introduces infinite higher derivative curvature terms
$R^{2}, { R }_{ \mu \nu  }{ R }^{ \mu \nu  }, R_{\mu\nu\kappa\lambda} 
R^{\mu\nu\kappa\lambda}\dots$. 
Although these higher derivative curvature terms are indispensable
to construct the self-consistent renormalizable theory of quantum gravity,
they lead to several pathologies.
We clearly show that even if they are written as the 
Planck-suppressed operators 
they lead to serious consequences and 
and de Sitter or radiation-dominated Universe is highly unstable.
We show that the couplings of these higher derivative curvatures 
must satisfy $\left|a_{1,2,3}\right|\gtrsim 10^{118}$
to be consistent with the cosmological observations.
Thus, the standard effective field theories of quantum gravity
fail to describe the observed Universe unless introducing a specific
technique dealing with the higher derivative curvature terms.\end{abstract}
\end{titlepage}

\newpage

%%%%%%%%%%%%%%%%%%%%%%%%%%%%%%%%%%%%%%%%%%%%%%%%
%%%%%%%%%%%%%%%%%%%%%%%%%%%%%%%%%%%%%%%%%%%%%%%%
\section{Introduction}
\label{sec:intro}
%%%%%%%%%%%%%%%%%%%%%%%%%%%%%%%%%%%%%%%%%%%%%%%%
%%%%%%%%%%%%%%%%%%%%%%%%%%%%%%%%%%%%%%%%%%%%%%%%
To construct quantum field theory (QFT) of gravity 
has many serious problems. Famously, general relativity 
is not renormalizable which is a reasonable requirement on fundamental theory.
Adding higher curvature terms
$R^{2}, { R }_{ \mu \nu  }{ R }^{ \mu \nu  }, R_{\mu\nu\kappa\lambda} 
R^{\mu\nu\kappa\lambda}$~\cite{Stelle:1976gc} to 
Einstein-Hilbert action makes the theory 
renormalizable or even superrenormalizable~\cite{Buchbinder:1992rb,Asorey:1996hz}.
However, they lead to unphysical massive 
ghosts and spacetime instabilities. 
The spin-$2$ massive ghost brings a notorious 
unitary problem about 
the gravitational $S$-matrix although
further higher-curvature corrections might save 
such a unitary problem~\cite{Tomboulis:1980bs,Antoniadis:1986tu,Modesto:2011kw}.
Furthermore, these higher curvatures destabilize 
classical spacetimes under small perturbations~\cite{Horowitz:1978fq,RandjbarDaemi:1981wd,
Jordan:1987wd,Suen:1989bg,Suen:1988uf,Salles:2014rua} and provide unstable 
de Sitter spacetime solutions~\cite{Starobinsky:1980te,Azuma:1986st}.
Although they are indispensable ingredients
for the renormalization~\cite{Stelle:1976gc} and 
constricting the perturbative effective field theories 
of quantum gravity (QG) or string theory~\cite{Zwiebach:1985uq,Deser:1986xr,Donoghue:1994dn},
they provide undesired pathology and 
there has been considerable debate about these issues.

In the perturbative effective field theory of gravity,
the higher curvature terms  
appear as quantum corrections 
in Einstein-Hilbert action and the effective action of the gravity 
can be given by~\cite{Donoghue:1994dn},
\begin{align}\label{eq:effective}
\Gamma_{\rm eff} \left[g_{\mu\nu}\right]&=  
-\frac { 1 }{ 16\pi { G }_{ N } } \int { d }^{ 4 }x\sqrt { -g } ( R+2\Lambda
+c_1R^2\nonumber \\&
+c_2R_{\mu\nu}R^{\mu\nu}+c_3R_{\mu\nu\kappa\lambda}
R^{\mu\nu\kappa\lambda}+c_4\Box R+\cdots  \biggr) \, ,
\end{align}
where the parameters of the effective action
as the cosmological constant $\Lambda$ and 
Newton's constant ${ G }_{ N }$ are 
determined by experiments and observations. 
The higher-derivative curvatures $R^{2}, 
{ R }_{ \mu \nu  }{ R }^{ \mu \nu  }, R_{\mu\nu\kappa\lambda} 
R^{\mu\nu\kappa\lambda}$ are leading quantum corrections 
and express gravitational vacuum polarization or quantum particle creation.
For the effective field theory approaches,
these coefficients are expected to be
$c_{1,2\dots}\sim \frac{\mathcal{N}}{M_{\rm P}^2}$
($\mathcal{N}$ is a particle number for theory)
and are strongly suppressed by Planck mass or reduced Planck mass
which is defined by $M_{\rm P}^{2}=1/8\pi G_{N}$.
Hence, one might consider that the low-energy physics and 
Planck-scale physics are safely sequestered. 
For instance, the Newtonian potential for 
gravitational interactions of two heavy 
objects can be described by the effective action~\cite{Donoghue:1993eb},
\begin{equation}
V\left( r \right) =-\frac { G_{ N }{ m }_{ 1 }{ m }_{ 2 } }{ r } 
\left[ 1-\frac { G_{ N }\left( { m }_{ 1 }+{ m }_{ 2 } \right)  }{ r } -\frac { 135 }{ 30{ \pi  }^{ 2 } } 
\frac { G_{ N }\hbar }{ { r }^{ 2 } } +\cdots \right] .
\end{equation}
where we explicitly denoted the Planck constant $\hbar$.
The first-order correction of $G_{ N }( { m }_{ 1 }+{ m }_{ 2 })/r$ 
comes from general relativity,
whereas the second-order correction expresses quantum corrections 
which are derived by $R^{2}, 
{ R }_{ \mu \nu  }{ R }^{ \mu \nu  }, R_{\mu\nu\kappa\lambda} 
R^{\mu\nu\kappa\lambda}$.
These quantum corrections 
are safely negligible for the Newtonian potential
and such quantum effects appear only at 
the very short distance.
The above argument matches the heart of 
the perturbative effective field theories which are
a standard paradigm of particle physics~\cite{Weinberg:1978kz}.
Hence, one expects that 
these quantum corrections are irrelevant in the low-energy physics
and our universe would not receive such quantum effects 
except for very early stage like 
singularities~\cite{Anderson:1983nq,Anderson:1984jf,Nojiri:2004ip}.

However, these higher-derivative curvature terms 
modify Einstein equations to higher-derivative equations~\cite{Ford:2005qz}
and, consequently, the gravitational system has well-known 
Ostrogradski instability~\cite{Woodard:2006nt}.
Even if such corrections are suppressed by the reduced Planck mass $M_{\rm P}$,
the spacetime drastically changes 
compared with general relativity.
Although these issues about the higher derivative quantum gravity 
or $f(R)$-gravity theories
have been discussed in Refs.~\cite{Arbuzova:2011fu,Myrzakulov:2014hca,
Nojiri:2010wj,Clifton:2011jh,Capozziello:2011et,Cusin:2015oza,Biswas:2016egy}, 
they have not been fully clarified and considered in 
perturbative effective field theories of QG.

The present paper addresses the problem involving effective field theories of QG
and discusses the spacetime instability induced by higher-derivative curvatures.
In principle, we focus on the effective field theory approach of QG 
originating from Einstein gravity, 
but our results are directly applied to the higher-derivatives gravity theories.
The present paper is organized as follows. 
In Section~\ref{sec:Renormalization}, we review the renormalization of the gravity 
why higher-derivative curvatures are indispensable for QG.
In Section~\ref{sec:instability}, we consider the effective action of the gravity and 
derive modified Einstein equations. 
Solving the modified Einstein equations, we investigate 
the instabilities of the Friedmann-Lemaitre-Robertson-Walker
(FLRW) spacetime. 
In our analysis, we assume that the initial background solution is the solution of the 
Einstein gravity with $H_i$, and discuss how much it varies on a timescale from that solution.
We show that the 
de Sitter spacetime is unstable 
against small perturbations
%%%%%%%%%%%%%%%
\footnote{
The similar instability of de Sitter spacetime from higher-derivative quantum gravity
has been discussed in~\cite{Myrzakulov:2014hca}.}
%%%%%%%%%%%%%%%
and the de Sitter expansion 
rolls down to the Planckian stage 
or terminates even in one normalization time $\tau=H_i\cdot t$ 
where $H_i$ is the initial value of the Hubble parameter,
$t$ is the cosmic time, and we introduce a dimensionless timescale 
parameter $\tau$ to simplify our discussion.
The instabilities are consistent with the early results of Refs.~\cite{Suen:1989bg,Suen:1988uf}.
The radiation dominated FLRW spacetime is also unstable and drastically changes
in one normalization time $\tau$.
We clearly show that the homogenous and isotropic flat Universe is unstable 
and either grow exponentially 
or oscillate even in Planckian time $t_{\rm I}=(\alpha_{ 1 }G_N)^{1/2}\approx 
\alpha_{ 1 }10^{-43}\ {\rm sec}$.
In Section~\ref{sec:conclusion} 
we draw the conclusion of our work.

%%%%%%%%%%%%%%%%%%%%%%%%%%%%%
%%%%%%%%%%%%%%%%%%%%%%%%%%%%%
\section{Renormalization and effective action for gravity}
\label{sec:Renormalization}
%%%%%%%%%%%%%%%%%%%%%%%%%%%%%
%%%%%%%%%%%%%%%%%%%%%%%%%%%%%
The effective field theory of quantum gravity is a full quantum theory 
and should take into account the loop diagrams of the various fields~\cite{Donoghue:1994dn}.
Accidentally, all quantum corrections of gravity are renormalized into 
$R^{2}, { R }_{ \mu \nu  }{ R }^{ \mu \nu  }, R_{\mu\nu\kappa\lambda} 
R^{\mu\nu\kappa\lambda}$ and the simplest 
renormalizable gravitational action can be contracted by using them.
The only difference between higher-derivative renormalizable gravity and
effective field theories of quantum gravity with respect to the leading 
quantum corrections is whether to introduce renormalized coupling constants or write down quantum corrections 
with ultraviolet (UV) cutoffs. Recalling that the Standard Model of particle theory is a renormalizable effective field theory, 
one can also treat the higher-derivative quantum gravity 
as an effective field theory and introduce an infinite higher-order curvature term. 
Hereafter, we will review how the higher-order curvature terms
incorporate the quantum effects of gravity.

We introduce the simplest renormalizable (bare)
gravitational action given as,
\begin{align}\label{eq:gravity}
S\left[g_{\mu\nu}\right] \equiv  -\frac { 1 }{ 16\pi { G }_{ N } } \int { { d }^{ 4 }x\sqrt { -g } \left( R+2\Lambda  \right)  }+{ S }_{ \rm HG }\left[g_{\mu\nu}\right]  +S_{\rm matter},
\end{align}
where ${ S }_{ \rm HG }$ is the higher-derivative gravitational action,
\begin{align}\label{eq:higher}
S_{\rm HG}\left[g_{\mu\nu}\right] =  \int { { d }^{ 4 }x\sqrt { -g } \left( { a }_{ 1 }R^2+{ a }_{ 2 }R_{\mu\nu}
R^{\mu\nu}
+{ a }_{ 3 }R_{\mu\nu\kappa\lambda} R^{\mu\nu\kappa\lambda}
+{ a }_{ 4 }\Box R \right)  },
\end{align}
with the higher-derivative curvature couplings $a_{1,2,3,\dots }$
and $S_{\rm matter}$ is the matter action.
These higher-derivative terms and couplings $a_{1,2,3,\dots }$ are indispensable
for the renormalization to eliminate one-loop divergences.
For instance, one-loop divergent corrections from scalar fields
are calculated by using Schwinger-DeWitt method and 
dimensional regularization as follows~\cite{Gorbar:2002pw}:
\begin{align}
&{ \Gamma  }^{ \left( \rm 1-loop \right)  }_{\rm eff}=\frac { -1 }{ 2{ \left( 4\pi  \right)  }^{ 2 }  } \int { { d }^{ 4  }x } \sqrt { -g } \Biggl\{ \left[ \ln { \left( \frac { { m }^{ 4 } }{ { \mu  }^{ 2 } }  \right)  } 
-\frac { 1 }{ \epsilon  } -\log { 4{ \pi  } } +\gamma+\cdots  \right] \times \biggl[\frac{1}{2}m^4+{ m }^{ 2 }{ \left( \xi -\frac { 1 }{ 6 }  \right)  }R\nonumber \\ & -\frac { 1 }{ 6 } \left( \xi -\frac { 1 }{ 6 }  \right) \Box R+\frac { 1 }{ 2 } { \left( \xi -\frac { 1 }{ 6 }  \right)  }^{ 2 }{ R }^{ 2 }+\frac { 1 }{ 180 } \left( R_{\mu\nu\kappa\lambda}
R^{\mu\nu\kappa\lambda}-R_{\mu\nu}
R^{\mu\nu}-\Box R \right) \biggr] \Biggr\},
\end{align}
where $\mu$ is the subtraction scale, 
$\epsilon$ is the regularization parameter and 
$\gamma$ is the Euler's constant, and 
$m$ or $\xi$ is the mass or non-minimal 
coupling of the scalar field.
The higher-derivative corrections written as 
divergent quantum corrections
express gravitational vacuum polarization 
or quantum particle creation.
These divergences can be absorbed by (bare) coupling constants 
of the gravitational action of Eq.~(\ref{eq:gravity}) and Eq.~(\ref{eq:higher}).
Hence, we get renormalized coupling constants.

Proceeding to the renormalization, for instance, 
we obtain the renormalized cosmological constant,
\begin{equation}
\frac { { \Lambda^{\rm ren}  } }{ 8\pi { G }_{ N }^{\, \rm ren}   }  =
\frac { { \Lambda \left(\mu\right) } }{ 8\pi { G_{ N } \left(\mu\right)}}+\frac { { m }^{ 4 } }{ 64{ \pi  }^{ 2 } } \left[ \ln { \left( \frac { { m }^{ 4 } }{ { \mu  }^{ 2 } }  \right)  } 
+ {\rm finite\ constant }  \right] ,
\end{equation}
which express physical cosmological constant and $\mu$ express the renormalization scale.
Recalling that the renormalized cosmological constant $\Lambda^{\rm ren}$
does not depend on the scale $\mu$,
we can get the renormalization
group equations for the cosmological constant,
\begin{align}
\mu \frac { d }{ d\mu  } \left( \frac { { \Lambda  } }{ 8\pi { G_{ N } } }  \left(\mu\right)\right) 
={ \beta  }_{ \Lambda  }=\frac { { m }^{ 4 } }{ 2{ \left( 4\pi  \right)  }^{ 2 } } ,
\end{align}
where $\beta_{ \Lambda  }$ is one-loop $\beta$-function for 
the cosmological constant. Similarly, we obtain the renormalization group equations
for other gravitational coupling constants.
If we consider $N_s$ real scalars with $m_s$, $N_f$ Dirac spinors with $m_f$
and $N_b$ massless vector bosons gravitational one-loop $\beta$-functions
are given as follows~\cite{Gorbar:2002pw,Shapiro:2008sf}, 
\begin{align}\begin{split}
&\mu \frac { d }{ d\mu  } \left( \frac { { \Lambda  } }
{ 8\pi { G_{ N } }}\left(\mu\right)  \right) 
={ \beta  }_{ \Lambda  }=\frac { N_{s} m _{s}^{ 4 } }{ 2{ \left( 4\pi  \right)  }^{ 2 } } 
- \frac { N_{f} m _{f}^{ 4 } }{ { \left( 4\pi  \right)  }^{ 2 } }  \\
&\mu \frac { d }{ d\mu  } \left( -\frac { { 1 } }{ 16\pi { G_{N} \left(\mu\right)  } }  \right) 
={ \beta  }_{ G_{N}  }
=\frac { N_{s} m _{s}^{ 2 } }{ { \left( 4\pi  \right)  }^{ 2 } } \left( \xi -\frac { 1 }{ 6 }  \right) 
+\frac {N_{f} m _{f}^{ 2 } }{ 3\, { \left( 4\pi  \right)  }^{ 2 }} \\
&\mu \frac { d{ a }_{ 1 } \left(\mu\right)  }{ d\mu  } ={ \beta  }_{ 1 }=
\frac { N_{s} }{ { 2\left( 4\pi  \right)  }^{ 2 } } { \left( \xi -\frac { 1 }{ 6 }  \right)  }^{ 2 }-
\frac {
5N_{f}+50N_{b} }{ { 360\left( 4\pi  \right)  }^{ 2 } },\quad
\mu \frac { d{ a }_{ 2 } \left(\mu\right)  }{ d\mu  } ={ \beta  }_{ 2 }=\frac { -N_{s}
+4N_{f}+88N_{b} }{ { 180\left( 4\pi  \right)  }^{ 2 } } \\
&\mu \frac { d{ a }_{ 3 } \left(\mu\right)  }{ d\mu  } ={ \beta  }_{ 3 }=\frac { 2N_{s}
+7N_{f}-26N_{b} }{ { 360\left( 4\pi  \right)  }^{ 2 } },\quad
\mu \frac { d{ a }_{ 4 } \left(\mu\right)  }{ d\mu  } ={ \beta  }_{ 4 }=\frac { N_{s}
+6N_{f}-18N_{b} }{ { 180\left( 4\pi  \right)  }^{ 2 } }
\end{split}\end{align}
where $a_{1,2,3,4}$ can not be fixed to 
be zero due the renormalization group (RG) running
and they are expected to be
\begin{align}
\frac { { \Lambda } }{ 8\pi { G }_{ N } } 
\sim \mathcal{N}{ \Lambda  }_{ \rm UV }^4,\quad
\frac { 1 }{ 16\pi { G }_{ N }  } 
\sim \mathcal{N}{ \Lambda  }_{ \rm UV }^2,\quad
{ a }_{ 1,2,3,4 }\sim \mathcal{N},
\end{align}
where ${ \Lambda  }_{ \rm UV }$ is the cut-off scale
and a large number of particle species $\mathcal{N}$ 
brings fine-tuning problems to the gravitational couplings.
Clearly, the cosmological constant ${ \Lambda }$ and the 
Newton's constant ${ G }_{ N }$ 
must permit a hard fine-tuning against the quantum corrections.
On the other hand, these higher-gravitational terms are interpreted as 
gravitational vacuum polarization or 
quantum particle production from gravity. 
Indeed, the particle creation ratio $p_{\, \rm creation}$
for the scalar field in the FLRW spacetime 
can be expressed by the higher-derivative terms~\cite{Dobado:1998mr},
\begin{align}
p_{\, \rm creation}\simeq 2 \cdot {\rm Im} \, { \Gamma  }_{\rm eff}^{ \left( \rm 1-loop \right)  }
& \simeq \frac{1}{16\pi^2}\int d^4x \sqrt{-g} 
\left(\frac{1}{180}R_{\mu\nu\kappa\lambda}
R^{\mu\nu\kappa\lambda}-\frac{1}{180}R_{\mu\nu}
R^{\mu\nu}\right.\nonumber\\
&+
\left.\frac{ 1 }{2}\left(\frac{1}{6}-\xi\right)^2 R^2\right)
+{\cal O}({R}^3),
\label{imea}
\end{align}
which is also consistent with the mode-mixing Bogolyubov technique.
Clearly, these terms can not be regarded as low-energy 
decoupling effects of the cosmological constant $\Lambda$ and 
the Newton's constant ${ G }_{ N }$ unlike the QED case
(see e.g. the detailed discussion in~\cite{Gorbar:2002pw}).
We emphasize that these higher curvature terms are indispensable
for the effective field theory of quantum gravity.

%%%%%%%%%%%%%%%%%%%%%%%%
%%%%%%%%%%%%%%%%%%%%%%%%
\section{Quantum gravitational instabilities of spacetime}
\label{sec:instability}
%%%%%%%%%%%%%%%%%%%%%%%%
%%%%%%%%%%%%%%%%%%%%%%%%

In this section, we will consider spacetime instability using the 
effective action~(\ref{eq:effective}) or
(\ref{eq:gravity}) and (\ref{eq:higher}) of the gravity at the one-loop level.
The higher-derivative terms modify the Einstein's equations
and destabilize classical solutions of the spacetime
even for the small perturbations. 
Here, we investigate the instabilities 
for the FLRW spacetime with various conditions 
and seek the stability condition.

The effective action of Eq.~(\ref{eq:gravity}) and (\ref{eq:higher}) 
derives the following modified Einstein's equations~\cite{Birrell:1982ix},
%%%%%%%%%%%%%%%%%%%%%%%%%%%%%%%%%%%
\footnote{The Gauss-Bonnet term reduces to a topological surface term in $D=3+1$ and 
we can neglect it in modified Einstein’s equations.}
%%%%%%%%%%%%%%%%%%%%%%%%%%%%%%%%%%%
\begin{align}\label{eq:classical}
\frac { 1 }{ 8\pi { G }_{ N } } \left( 
{ R }_{ \mu \nu  }-\frac { 1 }{ 2 }R{ g }_{ \mu \nu  }
+{ \Lambda  } { g }_{ \mu\nu }\right)
+{ a }_{ 1 }{ H }_{ \mu\nu }^{ \left( 1 \right)  }
+{ a }_{ 2 }{ H }_{ \mu\nu  }^{ \left( 2 \right)  }
+{ a }_{ 3 }{ H }_{ \mu\nu  }=\left< { T }_{ \mu \nu  } \right> ,
\end{align}
where $\left< { T }_{ \mu \nu  } \right>$ is the vacuum expectation value of 
the energy-momentum tensor and,
\begin{align*}
\begin{split}
H^{(1)}_{\mu\nu} &\equiv 
\frac { 1 }{ \sqrt { -g }  } \frac { \delta  }{ \delta { g }^{ \mu \nu  } } \int { d } ^{ 4 }x\sqrt { -g } { R }^{ 2 }=
2\nabla_\nu \nabla_\mu R -2g_{\mu\nu}\Box R
 - {1\over 2}g_{\mu\nu} R^2 +2R R_{\mu\nu},   \\
H^{(2)}_{\mu\nu} &\equiv
\frac { 1 }{ \sqrt { -g }  } \frac { \delta  }{ \delta { g }^{ \mu \nu  } } \int { d } ^{ 4 }x\sqrt { -g }
R_{\mu\nu}R^{\mu\nu}=
2\nabla_\alpha \nabla_\nu R_\mu^\alpha - \Box R_{\mu\nu} -{1\over 2}g_{\mu\nu}\Box R
  -{1\over 2}g_{\mu\nu} R_{\alpha\beta}R^{\alpha\beta} 
   +2R_\mu^\rho R_{\rho\nu},     \\
H_{\mu\nu} &\equiv 
\frac { 1 }{ \sqrt { -g }  } \frac { \delta  }{ \delta { g }^{ \mu \nu  } } \int { d } ^{ 4 }x\sqrt { -g }
R_{\mu\nu\kappa\lambda}R^{\mu\nu\kappa\lambda}=
-  H^{(1)}_{\mu\nu} + 4H^{(2)}_{\mu\nu}.
\end{split}
\end{align*}
Since left-hand side of Eq.~(\ref{eq:classical}) is 
covariantly conserved, the quantum energy-momentum tensor must 
satisfy covariant conservation: ${ \nabla  }^{ \mu  }\left< { T }_{ \mu \nu  } \right>=0$. 
For a flat FLRW universe, the geometrical tensors $H^{(1)}_{\mu\nu}$ and
$H^{(2)}_{\mu\nu}$ are related with $H^{(1)}_{\mu\nu}=3H^{(2)}_{\mu\nu}$.
Thus,  we obtain the following relation,
\begin{align}
{ a }_{ 1 }{ H }_{ \mu\nu }^{ \left( 1 \right)  }
+{ a }_{ 2 }{ H }_{ \mu\nu  }^{ \left( 2 \right)  }+{ a }_{ 3 }{ H }_{ \mu\nu  }=
\left({ a }_{ 1 }+\frac{1}{3}{ a }_{ 2 }+\frac{1}{3}{ a }_{ 3 }\right){ H }_{ \mu\nu }^{ \left( 1 \right)  }
={ \alpha }_{ 1 }{ H }_{ \mu\nu }^{ \left( 1 \right)  },
\end{align}
in which we introduce ${ \alpha }_{ 1 }={ a }_{ 1 }+\frac{1}{3}{ a }_{ 2 }+\frac{1}{3}{ a }_{ 3 }$.
However, the quantum energy-momentum tensor $\left< { T }_{ \mu \nu  } \right>$
requires more additional geometric tensors (for the detailed discussions 
see Ref~\cite{Birrell:1982ix}).
For instance, the renormalized vacuum energy-momentum tensor
for a massless conformal coupled scalar field 
which corresponds to the conformal anomaly~\cite{Capper:1974ic,Deser:1976yx,Duff:1977ay,Christensen:1978gi}, 
is given as follows,
\begin{align}\label{eq:conformal}
\left< { T }_{ \mu \nu  } \right>_{\rm conformal}=\frac{1}{2880\pi^2} 
\left(-\frac{1}{6}{ H }_{\mu \nu   }^{ \left( 1 \right)}
+ { H }_{ \mu \nu   }^{ \left( 3 \right)} \right) ,
\end{align}
where ${ H }_{ \mu \nu   }^{ \left( 3 \right)}$ is introduced by the conformal anomaly 
\begin{align*}
H^{(3)}_{\mu\nu} 
&\equiv  \frac{1}{12}R^2g_{\mu\nu}-R^{\rho\sigma}R_{\rho\mu\sigma\nu}  \\
&=R_\mu^\rho R_{\rho\nu} - \frac{2}{3}R R_{\mu\nu}
  -\frac{1}{2}R_{\rho\sigma}R^{\rho\sigma}g_{\mu\nu}  
  +\frac{1}{4}R^2g_{\mu\nu} ,     
\end{align*}
Furthermore, we must introduce an additional geometric tensor ${ H }_{ \mu \nu   }^{ \left( 4 \right)}$
which depends on the vacuum state like Eq.~(\ref{eq:Bunch-Davies})
(see Ref~\cite{Birrell:1982ix} for the details).
From here we drop the brackets of $\left< { T }_{ \mu \nu  } \right>$ and 
simply neglect ${ H }_{ \mu \nu   }^{ \left( 4 \right)}$. 
Hence, we obtain the following modified Einstein's equations taking account of 
quantum gravitational effects,
%%%%%%%%%%%%%%%%%%%%%%%%%%%%%%%%%%%%
\footnote{~
The parameter ${ \alpha }_{ 3 }$ for conformal invariant fields 
is expected to be~\cite{Birrell:1982ix}, 
\label{eq:alpha}
\begin{align}
{ \alpha }_{ 3} = \frac{1}{2880\pi^{2}}\left( N_{s}+\frac{11}{2}N_{f}+ 62 N_{b}\right), 
\end{align}}
%%%%%%%%%%%%%%%%%%%%%%%%%%%%%%%%%%%%
\begin{align}\label{eq:semiclassical}
\frac { 1 }{ 8\pi { G }_{ N } } \left( 
{ R }_{ \mu \nu  }-\frac { 1 }{ 2 }R{ g }_{ \mu \nu  }
+{ \Lambda  } { g }_{ \mu\nu }\right)
+{ \alpha }_{ 1 }{ H }_{ \mu\nu }^{ \left( 1 \right)  }
+{ \alpha }_{ 3 }{ H }_{ \mu\nu  }^{ \left( 3 \right)  }={ T }_{ \mu \nu  } ,
\end{align}

Thus, we can get a differential equation for the flat FLRW spacetime,
\begin{align}\label{eq:H-equation}
\begin{split}
\frac{\dot{a}^2}{a^2}&= \frac{\Lambda}{3}
-{8\pi G_N }\frac{18\alpha_1}{3}\left(2\frac { \dot { a } \dddot { a }  }{ { a }^{ 2 } } -\frac { { \ddot { a }  }^{ 2 } }{ { a }^{ 2 } } +2\frac { { \ddot { a } \dot { a }  }^{ 2 } }{ { a }^{ 3 } } -3\frac { \dot { a }^{ 4 }   }{ a^{ 4 }  }\right)\\
&+{8\pi G_N }\alpha_3\left(\frac{\dot{a}^4}{a^4}\right)
+ \frac{8\pi { G }_{ N }}{3}\rho_{\rm matter},
\end{split}
\end{align}
Then, we rewrite Eq.~(\ref{eq:H-equation}) concerning the Hubble parameter
%%%%%%%%%%%%%%%%%%%%%%%%%%%%%%%%%%%%%%%%%%%%%%%%
%%%%%%%%%%%%%%%%%%%%%%%%%%%%%%%%%%%%%%%%%%%%%%%%
\footnote{~
The derivative of Eq.~(\ref{eq:Hubble-equation}) yields, 
\begin{align*}
\begin{split}
\dot { H }&={16\pi G_N }\alpha_3{ H }^{ 2 }\dot { H }
-{48\pi G_N }{\alpha_1}\\
&\times \left(6\dot { H }^{ 2 } +3H\ddot { H } + \dddot { H }\right)
-4\pi { G }_{ N }\left( 1+\omega \right)\rho_{\rm matter}  \,.
\end{split}
\end{align*}
which includes the covariant conservation law of Eq.~(\ref{eq:conservation-law}).},
%%%%%%%%%%%%%%%%%%%%%%%%%%%%%%%%%%%%%%%%%%%%%%%%
%%%%%%%%%%%%%%%%%%%%%%%%%%%%%%%%%%%%%%%%%%%%%%%%
\begin{align}\label{eq:Hubble-equation}
\begin{split}
{H^2}&= \frac{\Lambda}{3}
-{48\pi G_N }{\alpha_1}\left(6{ H }^{ 2 }\dot { H } +2H\ddot { H } -{ \dot { H }  }^{ 2 }\right)\\
&+{8\pi G_N }\alpha_3H^4+ \frac{8\pi { G }_{ N }}{3}\rho_{\rm matter},
\end{split}
\end{align}
where the energy density of matter satisfies the covariant conservation law,
\begin{align}\label{eq:conservation-law}
\dot {\rho}_{\rm matter}=-3H\left( \rho_{\rm matter}+P_{\rm matter}\right)=-
3H\left( 1+\omega \right)\rho_{\rm matter},
\end{align}
in which $w=P/\rho$ is an equation-of-state parameter.
For non-relativistic, relativistic matter or vacuum state we get 
$w = 0,1/3,-1$ respectively. 
Note that Einstein's equations have no additional terms and 
de Sitter spacetime is defined to be the vacuum spacetime: $H^2 = \Lambda/3$.

%%%%%%%%%%%%%%%%%%%%%%%%%%%%%%%%%
\subsection{De Sitter spacetime solutions from quantum corrections}
%%%%%%%%%%%%%%%%%%%%%%%%%%%%%%%%%
The modified Einstein's equations
are formally written as the higher-derivative equations, and therefore, 
they do not necessarily follow the standard description of general relativity.
We note that the vacuum state $w = -1$ of the effective Einstein's equations leads to two 
classical and quantum de Sitter solutions~\cite{Starobinsky:1980te}.
Discarding time-derivative terms of Eq.~(\ref{eq:Hubble-equation}),
we get stationary solutions for the Hubble parameter as follows:
\begin{align}\label{eq:Asymptote}
H^2=\left(\frac{1}{{16\pi G_N }\alpha_3} \right)\pm \frac{1}{\alpha_3}
\sqrt{\left(\frac{1}{16\pi G_N } \right)^2 +
\frac{\Lambda\alpha_3}{24\pi G_N }} \, .
\end{align}
For $\alpha_3>0$ and relatively small cosmological constant
$M_{\rm P}^2 \gg\frac{4\alpha_3\Lambda}{3} $,
we can get two de Sitter spacetime solutions~\cite{Starobinsky:1980te,Hawking:2000bb,
Shapiro:2001rh,Pelinson:2002ef,Nojiri:2003vn,Netto:2015cba},
\begin{align}
H_{\mathrm{C}} \simeq \sqrt{\frac{\Lambda}{3}},
\quad H_{\mathrm{Q}} \simeq\sqrt{\frac{1}{{8\pi G_N }\alpha_3}},
\end{align}
where $H_{\mathrm{C}}$ turns out to be a classical de Sitter solution which is the same as 
the general relativity and $H_{\mathrm{Q}}$ is quantum driven de Sitter  
solution.
On the other hand, the gravity theory has no 
quantum de Sitter solutions for $\alpha_3 < 0$.
These two de Sitter solutions are generally unstable for the small perturbation~\cite{Matsui:2018iez} and then 
other spacetime derived from Eq.~(\ref{eq:semiclassical})
show the instability. We consider this case in the next subsection.

%%%%%%%%%%%%%%%%%%%%%%%%%%%%%%%%%
\subsection{Analytical estimation for FLRW spacetime instability}
%%%%%%%%%%%%%%%%%%%%%%%%%%%%%%%%%

We briefly discuss why the modified Einstein's equations~(\ref{eq:Hubble-equation})
provide instability.
Let us rewrite Eq.~(\ref{eq:Hubble-equation}) as follows, 
\begin{align}
\begin{split}
6{ H }^{ 2 }\dot { H } +2H\ddot { H } -{ \dot { H }  }^{ 2 }
= -\frac{H^2}{{48\pi G_N }{\alpha_1}}+
\frac{\Lambda}{{144\pi G_N }{\alpha_1}}
+\frac{\alpha_3}{6\alpha_1}H^4+ \frac{\rho_{\rm matter}}{18\alpha_1}.
\end{split}
\end{align}
For the smallness of $G_N\alpha_1$ we can approximately get the 
following equations 
\begin{align}\label{eq:ht-equation}
\begin{split}
\frac{d^2H}{dt^2}\approx-\frac{H}{96\pi G_N\alpha_1}.
\end{split}
\end{align}
This admits that the solutions either grow exponentially 
or oscillate even in Planckian time $t_{\rm P}=(96\pi\alpha_{ 1 }G_N)^{1/2}\approx 
\alpha_{ 1 }10^{-43}\ {\rm sec}$. 
For $\alpha_{ 1 }>0$ 
the perturbations oscillate in the Planck time
and they emit the Planck energy photons, $E\sim 10^{19}\ {\rm GeV}$~\cite{Horowitz:1978fq}
which is unreasonable for the observed Universe.
For $\alpha_{ 1 }<0$ the evolution of the Hubble parameter 
exponentially grow even in the Planck time and is not consistent with the observations.
Hence, the large values of the gravitational curvature coupling 
are required to stabilize 
the homogenous and isotropic flat Universe.

%%%%%%%%%%%%%%%%%%%%%%%%%%%%%%%%%
\subsection{Numerical estimation for FLRW spacetime instability}
%%%%%%%%%%%%%%%%%%%%%%%%%%%%%%%%%
Next, we numerically show the instability of the FLRW spacetime 
using Eq.~(\ref{eq:Hubble-equation}) and~(\ref{eq:conservation-law}).
First, let us start the de Sitter spacetime under small perturbations 
and consider the effects of gravity only. 
We rewrite Eq.~(\ref{eq:Hubble-equation}) and~(\ref{eq:conservation-law}) 
in terms of dimensionless quantities and obtain the 
following differential equation~\cite{Arbuzova:2011fu},
\begin{align}\label{eq:thubble}
\begin{split}
&h^{2}=-{x}h^{4}-y\left(6h^{2}h'+2hh''-h'^{2} \right)+z,\\
&z'=-
3h\left( 1+\omega \right)z\,. \end{split}
\end{align}
where we introduce $\tau =H_i t$, $h=H/H_i$, 
$x=- {8\pi G_N }\alpha_3{ H }^{ 2 }_i$,
$y={48\pi G_N }\alpha_1{ H }^{ 2 }_i$,
$z= \Lambda/3{ H }^{ 2 }_i+{8\pi G_N }\rho_{\rm matter}/3{ H }^{ 2 }_i$ 
and $H_i$ is the initial Hubble parameter at some time $t_i$.
For instance, we can expect the following cosmological relations,
\begin{align}
\begin{split}
&H_i \sim 10^{14 }\, {\rm GeV}, \ M_{\rm P} \sim 10^{18}\, {\rm GeV}, \ \alpha_{1,3}\sim 10^{-2}
\ \Longrightarrow \ x,y\sim 10^{-10}\\
&H_i \sim 10^{-42 }\, {\rm GeV}, \ M_{\rm P} \sim 10^{18}\, {\rm GeV}, \ \alpha_{1,3}\sim 10^{-2}
\ \Longrightarrow \ x,y\sim 10^{-122}
\end{split}
\end{align}
where the former corresponds to the typical Hubble parameter during inflation
from the current bound~\cite{Aghanim:2018eyx}
and the latter is consistent with the current Hubble parameter dominated by the dark energy.
These values for $\alpha_{1,3}$ 
are expected by the one-loop $\beta$-functions of high-derivative curvature couplings.
The dynamics of the dimensionless Hubble parameter $h$ with the vacuum state $w = -1$
and $\rho_{\rm matter}=0$
is determined by the following equation,
\begin{align}\label{eq:hubble}
h^{2}=-{x}h^{4}-y\left(6h^{2}h'+2hh''-h'^{2} \right)+z\,.
\end{align}
where prime express the derivative with respect to dimensionless time $\tau$.
The natural de Sitter initial conditions are given by
\begin{equation}
\tau_i=1,\quad h_i=1,\quad h_i'=0,\quad  z_i=1.
\end{equation}
which makes the time-evolution of the system consistent with the general relativity 
at the initial time $\tau_i$.

%%%%%%%%%%%%%%%%%%%%%%%%%%%%%%%%%%%%%%%%%%%%%%%%
%%%%%%%%%%%%%%%%%%%%%%%%%%%%%%%%%%%%%%%%%%%%%%%%
\begin{figure*}[t]
        \begin{tabular}{cc}
	\begin{minipage}{0.5\hsize}
		\centering
		\subfigure[$h_i=1+0.9$]{
		\includegraphics[width=75mm]{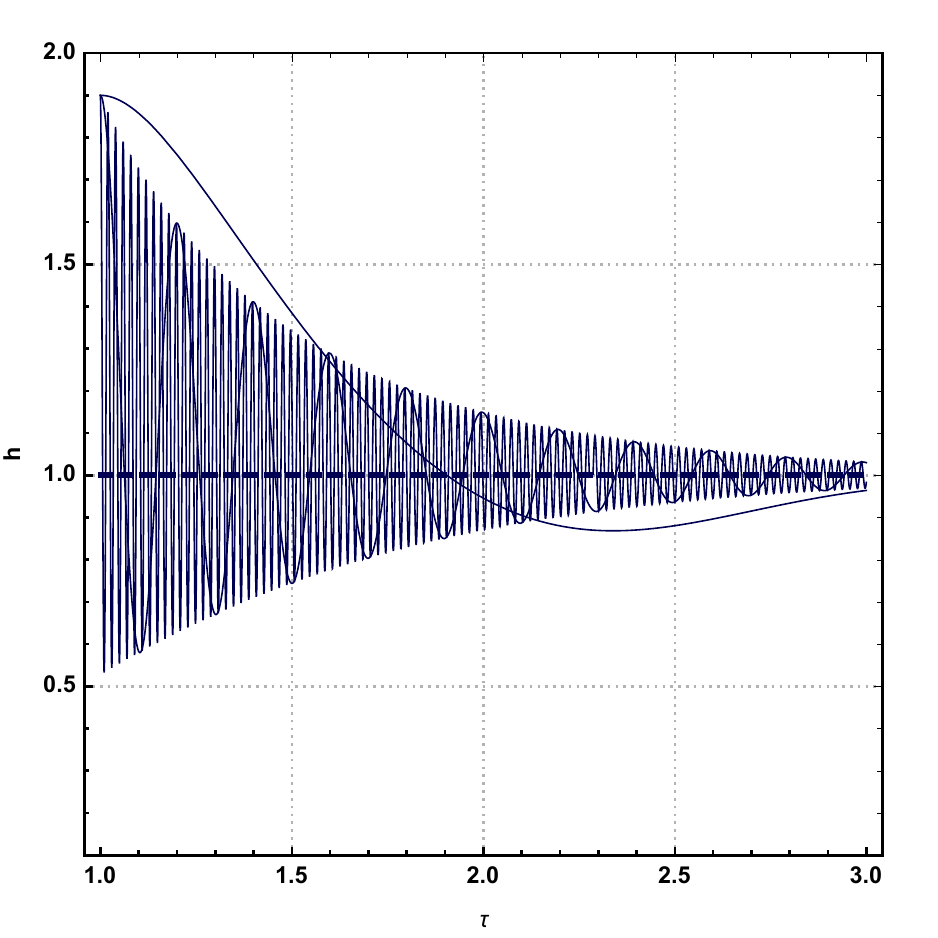}
		\label{fig:stabledeSitter1}}\end{minipage}
\begin{minipage}{0.5\hsize}
		\centering
		\subfigure[$h_i=1+0.1$]{
		\includegraphics[width=75mm]{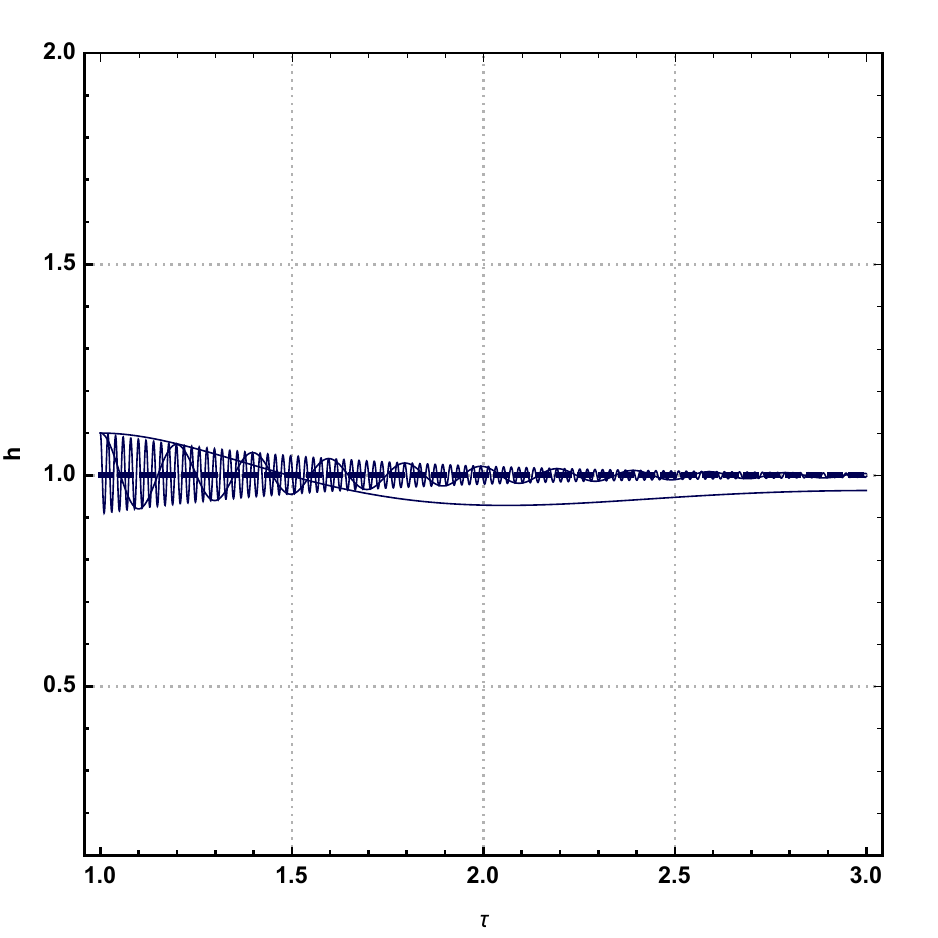}
		\label{fig:stabledeSitter2}}
\end{minipage}
\end{tabular}
\caption{For $y>0$, numerical solution of Eq.~(\ref{eq:hubble}) with the de Sitter initial conditions and the 
higher-derivative couplings of Eq.~(\ref{eq:deSitter1-initial}).
These figures show that the dynamics of the dimensionless Hubble parameter $h( \tau)$ in a few normalization time $\tau$.
The dashed line shows the usual de Sitter solution $h( \tau)=1$ from the general relativity.}
\end{figure*}
%%%%%%%%%%%%%%%%%%%%%%%%%%%%%%%%%%%%%%%%%%%%%%%%
%%%%%%%%%%%%%%%%%%%%%%%%%%%%%%%%%%%%%%%%%%%%%%%%

\begin{figure*}[t]
        \begin{tabular}{cc}
	\begin{minipage}{0.5\hsize}
		\centering
		\subfigure[$y=-10^{-1.0,-1.3,-1.5,-1.7,-1.9,-2.1}$]{
		\includegraphics[width=75mm]{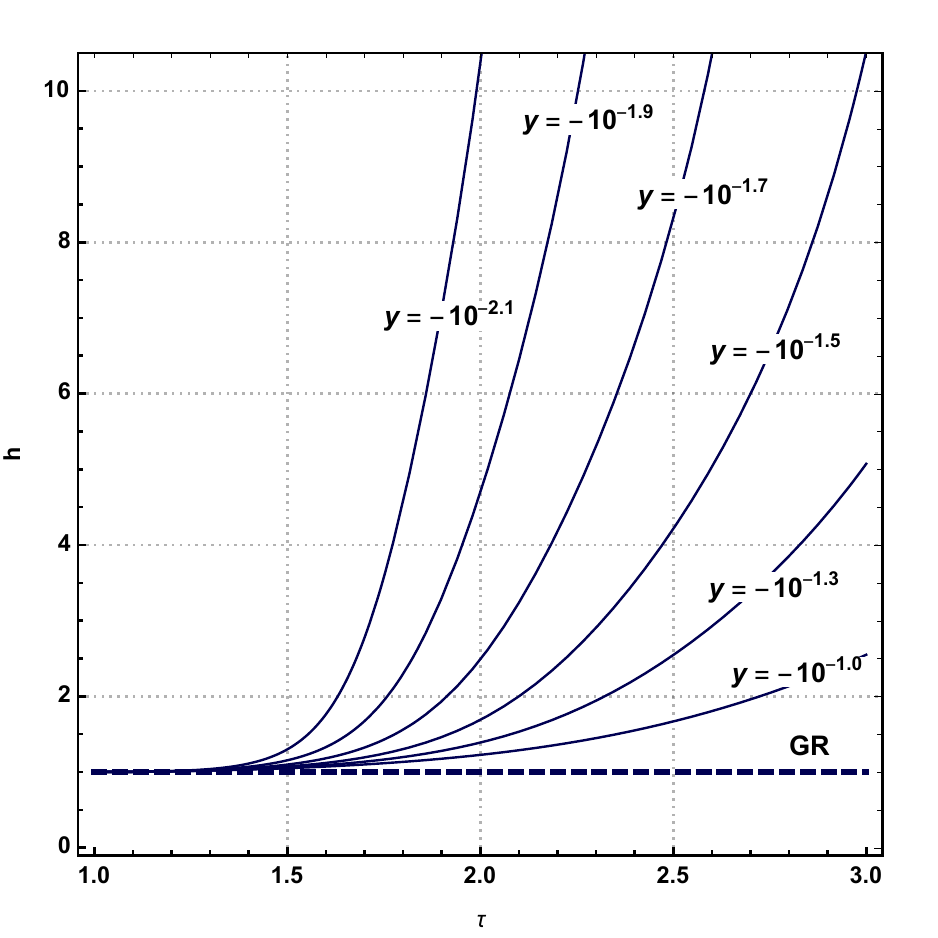}
		\label{fig:unstabledeSitter1}}\end{minipage}
\begin{minipage}{0.5\hsize}
		\centering
		\subfigure[$y=-10^{-9.8,-10.0,-10.2}$]{
		\includegraphics[width=75mm]{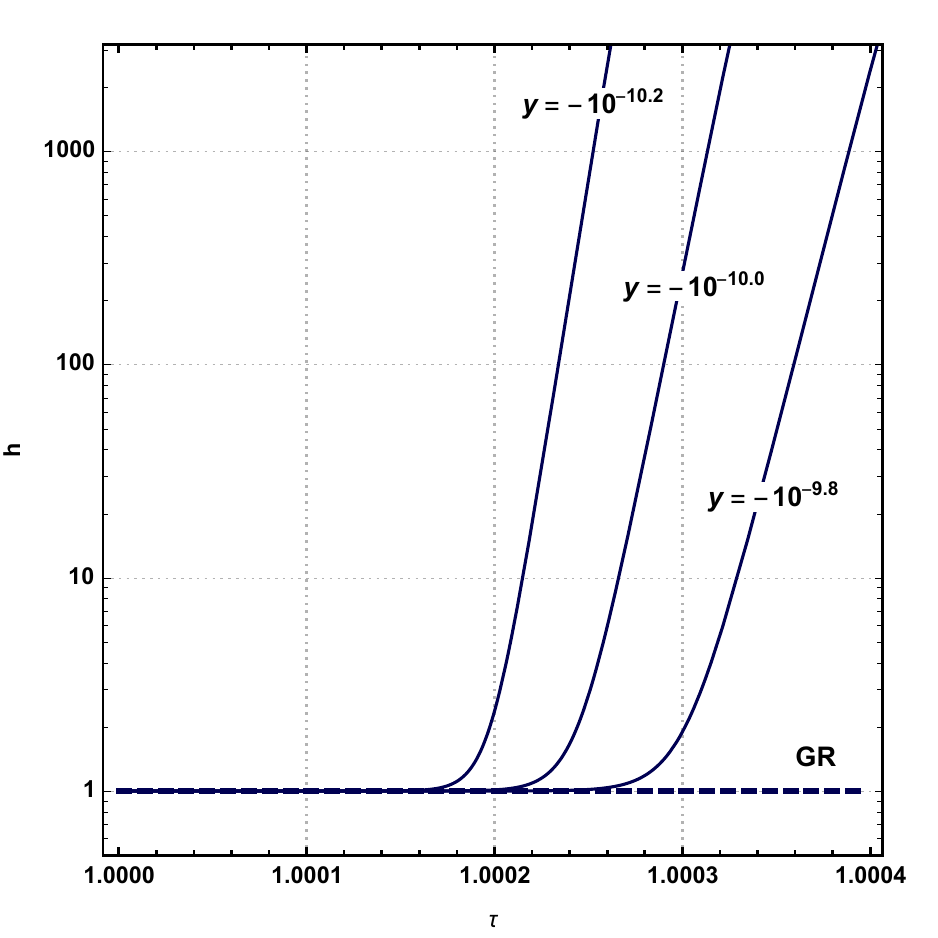}
		\label{fig:unstabledeSitter2}}
\end{minipage}
\end{tabular}
\caption{For $y<0$, numerical solution of Eq.~(\ref{eq:hubble}) with the de Sitter initial conditions and the 
higher-derivative couplings of Eq.~(\ref{eq:deSitter2-initial}). 
These figures show the instability for the dimensionless Hubble parameter $h( \tau)$ in a few normalization time $\tau$.
We demonstrate that the small values of $x,y$ amplify the spacetime instability.}
\end{figure*}

\begin{figure*}[t]
        \begin{tabular}{cc}
	\begin{minipage}{0.5\hsize}
		\centering
		\subfigure[$y=-10^{-1.0,-1.1,-1.2,-1.3,-1.4,-1.5}$ ]{
		\includegraphics[width=75mm]{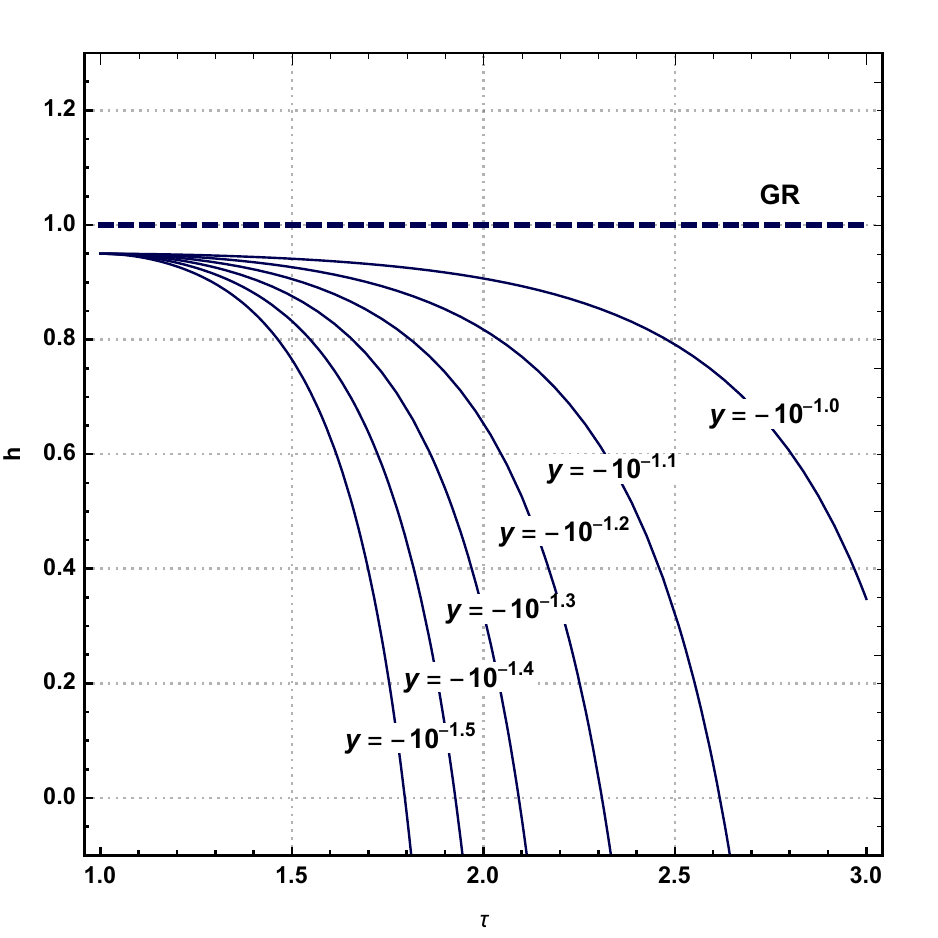}
		\label{fig:unstabledeSitter3}}\end{minipage}
\begin{minipage}{0.5\hsize}
		\centering
		\subfigure[$y=-10^{-10.0,-10.3,-10.6,-10.9}$]{
		\includegraphics[width=75mm]{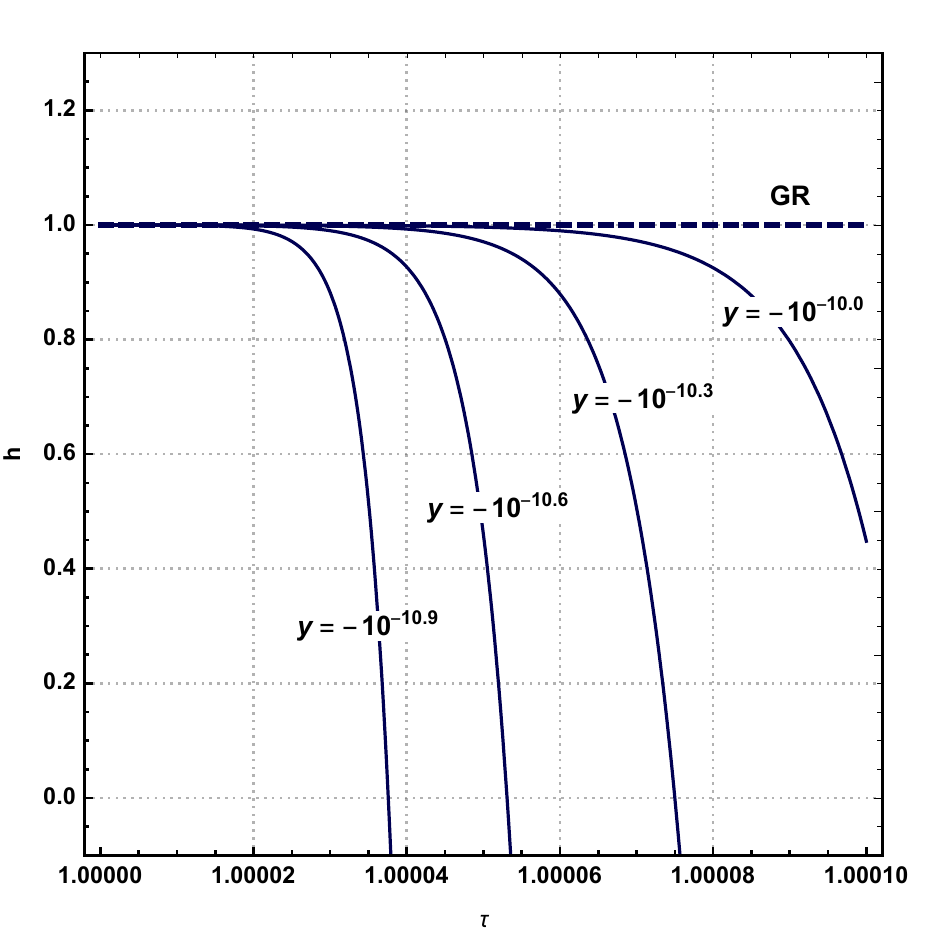}
		\label{fig:unstabledeSitter4}}
\end{minipage}
\end{tabular}
\caption{We show that 
numerical solutions of Eq.~(\ref{eq:hubble}) with the de Sitter conditions and the 
higher-derivative couplings of Eq.~(\ref{eq:deSitter2-initial}). 
These figures show the instability for dimensionless Hubble parameter $h( \tau)$ in a few normalization time $\tau$.
We show that for $h_{i}<1$ the de Sitter expansion terminates.}
\end{figure*}

We investigate the system of equations starting at $\tau_i=1$ with various conditions and perturbations. 
We find out that the numerical solutions of the system of equations
show the stability or instability which can be roughly understood by some analytical estimates.
In Fig.\ref{fig:stabledeSitter1} and \ref{fig:stabledeSitter2},
we present the numerical results for the dimensionless Hubble parameter $h$ determined from 
Eq.~(\ref{eq:hubble}) with the following initial conditions,
\begin{align}
\begin{split}\label{eq:deSitter1-initial}
&\textrm{Fig.\ref{fig:stabledeSitter1}:}\ h_i=1+0.9,\  h_i'=0,
\ x=10^{-1,-3,-5},\ y=10^{-1,-3,-5},\\
&\textrm{Fig.\ref{fig:stabledeSitter2}:}\ h_i=1+0.1,\  h_i'=0,
\ x=10^{-1,-3,-5},\ y=10^{-1,-3,-5},
\end{split}
\end{align}
and we compare them with the de Sitter solution 
$h( \tau)=1$ from the general relativity.
Fig.\ref{fig:stabledeSitter1} and \ref{fig:stabledeSitter2} show that the de Sitter spacetime oscillates for the perturbation and the variation converges 
for a few normalization times $\tau$. 
We found that the Hubble oscillation becomes faster for the small values of $x,y$
and the spacetime dynamics for $x,y>0$ or $x<0,\, y>0$
shows the same results.
In Fig.\ref{fig:unstabledeSitter1}, \ref{fig:unstabledeSitter2}, 
\ref{fig:unstabledeSitter3} and \ref{fig:unstabledeSitter4}, we show 
the numerical results for the dynamics of the dimensionless Hubble parameter $h$ 
for $y<0$ and take the following conditions,
\begin{align}
\begin{split}\label{eq:deSitter2-initial}
&\textrm{Fig.\ref{fig:unstabledeSitter1}:}\ h_i=1,\  h_i'=0,
\ x=10^{-1.0,-1.3,-1.5,-1.7,-1.9,-2.1},\ y=-10^{-1.0,-1.3,-1.5,-1.7,-1.9,-2.1},\\
&\textrm{Fig.\ref{fig:unstabledeSitter2}:}\ h_i=1,\  h_i'=0,
\ x=10^{-9.8,-10.0,-10.2},\ y=-10^{-9.8,-10.0,-10.2},\\
&\textrm{Fig.\ref{fig:unstabledeSitter3}:}\ h_i=1-10^{-1.3},\ h_i'=0,
\ x=10^{-1.0,-1.1,-1.2,-1.3,-1.4,-1.5},\ y=-10^{-1.0,-1.1,-1.2,-1.3,-1.4,-1.5},\\
&\textrm{Fig.\ref{fig:unstabledeSitter4}:}\ h_i=1-10^{-4.3},\ h_i'=0,
\ x=10^{-10.0,-10.3,-10.6,-10.9},\ y=-10^{-10.0,-10.3,-10.6,-10.9},
\end{split}
\end{align}
We found that the de Sitter spacetime is destabilized by
the Planck-suppressed quantum corrections in $\tau\sim \mathcal{O}(1)$
even if we set the tiny values of $x,y$. Rather, the smallness of $x,y$ amplifies the 
spacetime instability and this case is inconsistent with the usual general relativity.
In other words, the de Sitter spacetime is highly unstable for $\left| y \right| \ll 1$,
whereas the instability can be alleviated for $\left| y \right| \approx 1$.
This means a serious UV/IR mixing problem
and the higher-derivative curvatures can not be ignored unless 
the higher-derivative curvature couplings for the gravitational action 
are extremely large.

%%%%%%%%%%%%%%%%%%%%%%%%%%%%%%%%%%%%%%%%%%%%%%%%
%%%%%%%%%%%%%%%%%%%%%%%%%%%%%%%%%%%%%%%%%%%%%%%%
%%%%%%%%%%%%%%%%%%%%%%%%%%%%%%%%%%%%%%%%%%%%%%%%
%%%%%%%%%%%%%%%%%%%%%%%%%%%%%%%%%%%%%%%%%%%%%%%%
%%%%%%%%%%%%%%%%%%%%%%%%%%%%%%%%%%%%%%%%%%%%%%%%
%%%%%%%%%%%%%%%%%%%%%%%%%%%%%%%%%%%%%%%%%%%%%%%%

%%%%%%%%%%%%%%%%%%%%%%%%%%%%%%%%%%%%%%%%%%%%%%%%
%%%%%%%%%%%%%%%%%%%%%%%%%%%%%%%%%%%%%%%%%%%%%%%%
\begin{figure*}[t]
        \begin{tabular}{cc}
	\begin{minipage}{0.5\hsize}
		\centering
		\subfigure[$y=10^{-1.0,-3.0,-5.0}$]{
		\includegraphics[width=75mm]{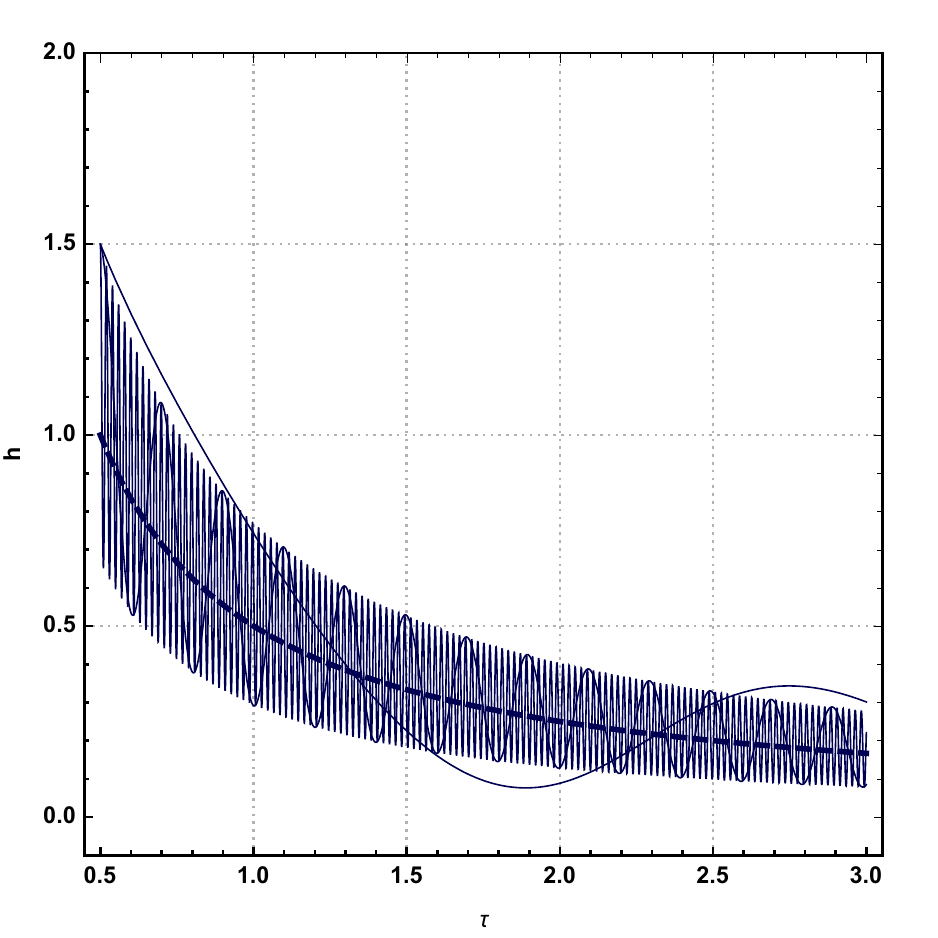}
		\label{fig:radiation1}}\end{minipage}
\begin{minipage}{0.5\hsize}
		\centering
		\subfigure[$y=-10^{-1.0,-1.3,-1.6,-1.9,-2.2}$]{
		\includegraphics[width=75mm]{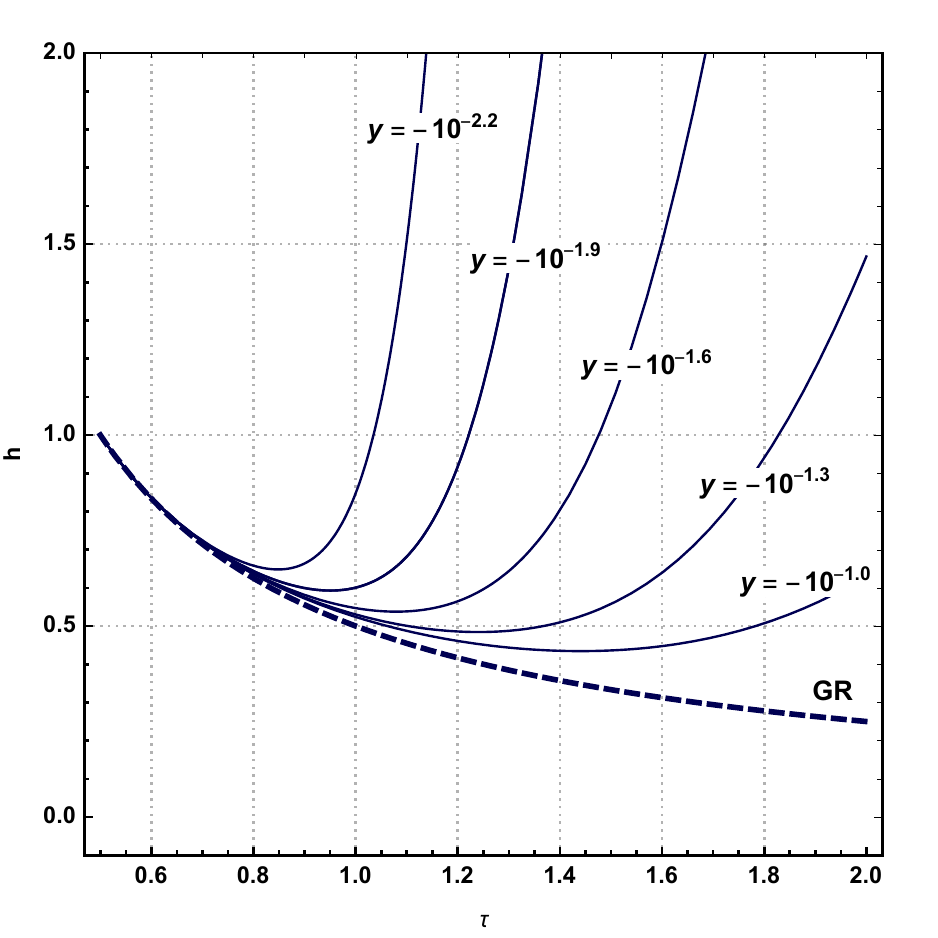}
		\label{fig:radiation2}}
\end{minipage}
\end{tabular}
\caption{We compare the numerical solution of Eq.~(\ref{eq:hubble}) with the conditions of Eq.~(\ref{eq:radiation-initial})
and the standard solution $h( \tau)=1/2\cdot\tau$ form the general relativity. 
Fig.\ref{fig:radiation1} show that de Sitter spacetime oscillates under the Hubble perturbations. 
Fig.\ref{fig:radiation2} show the instability for radiation-dominated Universe
and the solutions do not follow general relativity.  }
\end{figure*}
%%%%%%%%%%%%%%%%%%%%%%%%%%%%%%%%%%%%%%%%%%%%%%%%
%%%%%%%%%%%%%%%%%%%%%%%%%%%%%%%%%%%%%%%%%%%%%%%%
\begin{figure*}[h]
        \begin{tabular}{cc}
	\begin{minipage}{0.5\hsize}
		\centering
		\subfigure[$y=10^{-1.0,-3.0,-5.0}$ ]{
		\includegraphics[width=75mm]{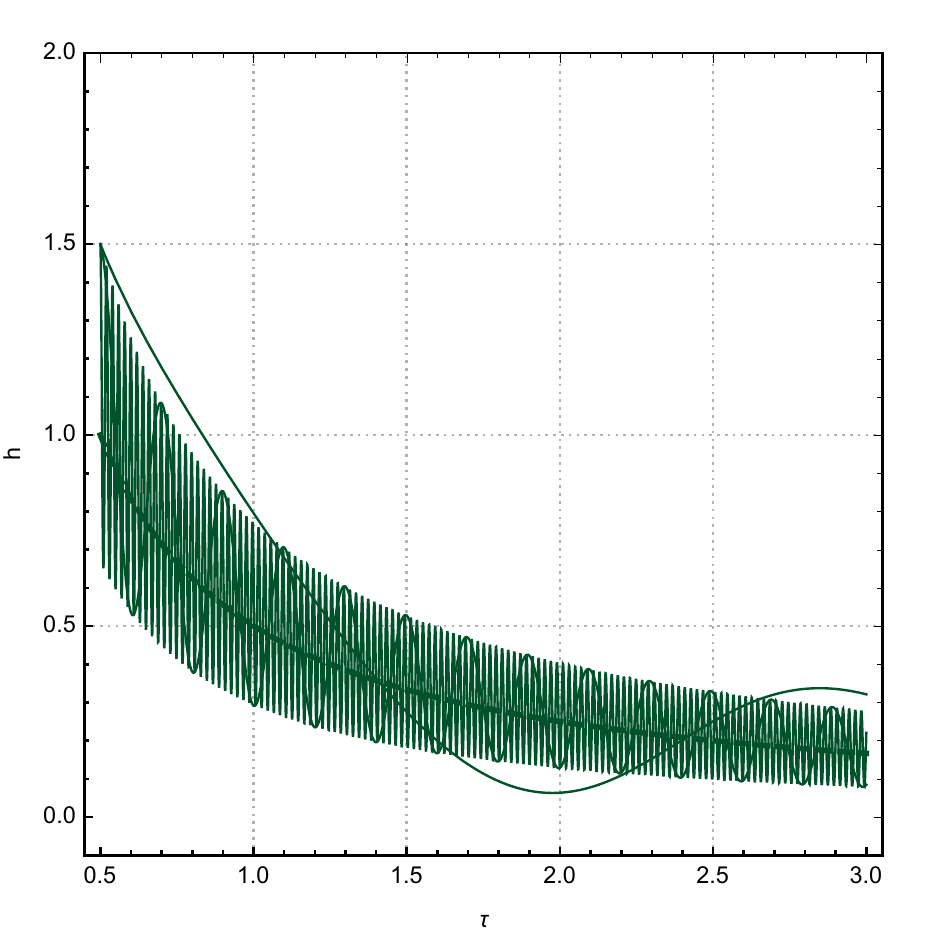}
		\label{fig:radiation3}}\end{minipage}
\begin{minipage}{0.5\hsize}
		\centering
		\subfigure[$y=-10^{-1.0,-2.0,-3.0}$]{
		\includegraphics[width=75mm]{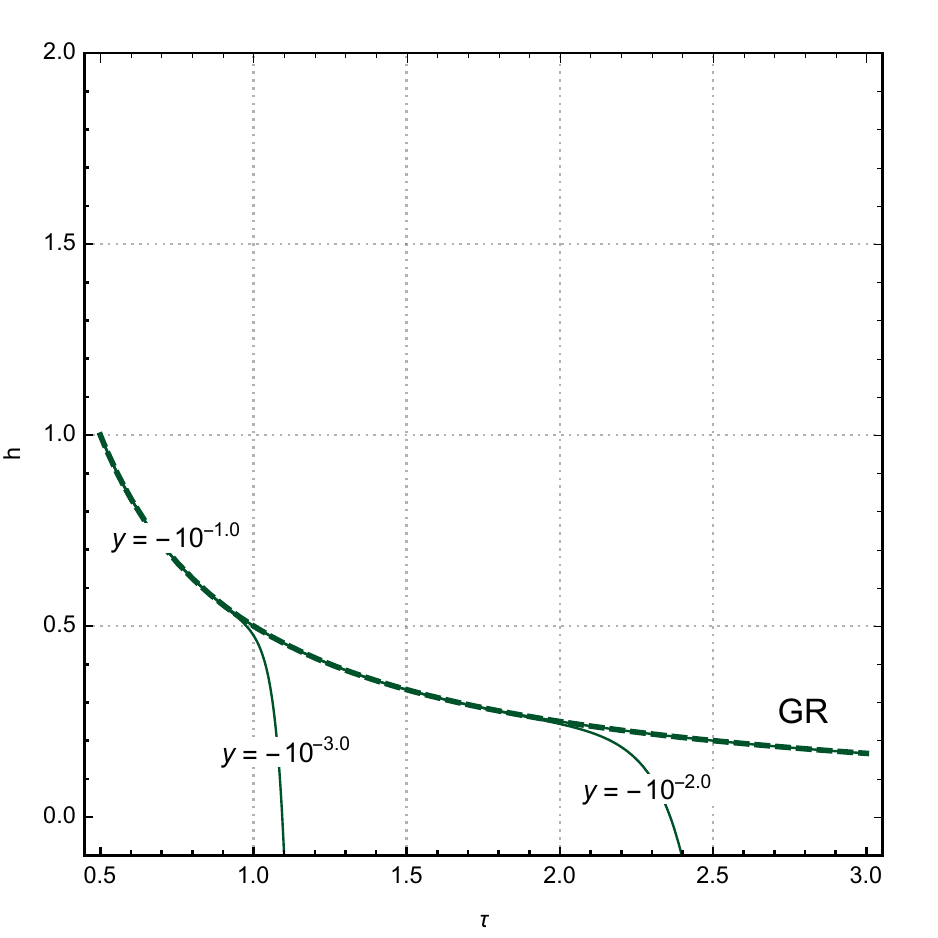}
		\label{fig:radiation4}}
\end{minipage}
\end{tabular}
\caption{Numerical solution of Eq.~(\ref{eq:hubble}) with the conditions of Eq.~(\ref{eq:radiation-initial})
where we set $x=0$ which is $R^2$-gravity. These figures show the instability for 
the dimensionless Hubble parameter $h( \tau)$ in a few normalization time $\tau$.
For $y>0$, the dynamics of the normalized Hubble parameter $h( \tau)$ is different 
from Fig.\ref{fig:radiation2}.
}
\end{figure*}
%%%%%%%%%%%%%%%%%%%%%%%%%%%%%%%%%%%%%%%%%%%%%%%%
%%%%%%%%%%%%%%%%%%%%%%%%%%%%%%%%%%%%%%%%%%%%%%%%

Next, we investigate the radiation-dominated Universe with 
the relativistic state $w = 1/3$.
For the radiation-dominated Universe,
we get $a\propto t^{1/2}$, $H\propto 1/(2 t)$.
Taking $h_i= 1$ and $h_i= 1/(2\tau_i)$ we can obtain  
the condition,
\begin{equation}
\tau_i=1/2,\quad h_i=1,\quad h_i'=-2,\quad z_i=1.
\end{equation}
where we rewrite $z= {8\pi G_N }\rho_{\rm matter}/3{ H }^{ 2 }_i$ and take $\Lambda=0$.
The condition is natural when we start the system 
at the radiation-dominated stage.
In Fig.\ref{fig:radiation1} and \ref{fig:radiation2} 
we investigate the system of equations starting at $\tau_i=1/2$  
by using Eq.~(\ref{eq:thubble}) with the following 
conditions and the higher-derivative parameters,
\begin{align}
\begin{split}\label{eq:radiation-initial}
&\textrm{Fig.\ref{fig:radiation1}:}\ h_i=1+0.5,\  h_i'=-2,
\ x=10^{-1,-3,-5},\ y=10^{-1,-3,-5},\\
&\textrm{Fig.\ref{fig:radiation2}:}\ h_i=1,\  h_i'=-2,
\ x=10^{-1.0,-1.3,-1.6,-1.9,-2.2},\ y=-10^{-1.0,-1.3,-1.6,-1.9,-2.2},\\
&\textrm{Fig.\ref{fig:radiation3}:}\ h_i=1+0.5,\  h_i'=-2,
\ x=0,\ y=10^{-1,-3,-5},\\
&\textrm{Fig.\ref{fig:radiation4}:}\ h_i=1,\  h_i'=-2,
\ x=0,\ y=10^{-1.0,-2.0,-3.0},
\end{split}
\end{align}
and compare them with the radiation-dominated
solution $h( \tau)=1/2\cdot\tau$.
Fig.\ref{fig:radiation1} show that the de Sitter spacetime oscillates for $y>0$ 
under the Hubble perturbations and the variations converge
to the solutions of the general relativity. In this case we found that 
the Hubble oscillations are faster for the small values of $x,y$.
On the other hand, Fig.\ref{fig:radiation2} show that
the higher-derivative curvature corrections lead to the instability
and for $y<0$ the solutions do not follow general relativity. 
For $x=0$, we demonstrate the dynamics of the normalized Hubble parameter $h( \tau)$
in Fig.\ref{fig:radiation3} and Fig.\ref{fig:radiation4}.

As we saw in the de Sitter spacetime
the smallness of $x,y$ amplifies the instability~\cite{Suen:1989bg,Suen:1988uf}.
Therefore, the Planck-suppressed curvature corrections strongly affect 
the spacetime dynamics. 
For $y< 0$, the higher-derivative curvature corrections 
clearly destabilize the classical spacetime and the 
universe at least has to satisfy the following conditions,
\begin{align}\label{eq:stable-condition}
y\left( \mu\right) >0 \ \Longrightarrow \ 
{8\pi G_N\left( \mu\right) }\left(\frac{18\alpha_1\left( \mu\right)}{3}\right)>0 ,
\end{align}
where $\mu$ is the renormalization scale. 
However, it is not easy to satisfy that condition
because the higher-derivative curvature couplings usually change sign
for the cosmological scale based on the renormalized group equations~\cite{Shapiro:2008sf}.
Hard fine-tuning is required for the couplings from the inflation to
the current accelerating stage 
and moreover, the above condition is only effective for the one-loop perturbative corrections.
Any higher-loop corrections require such conditions and that is not desired. 
Furthermore, for $y>0$ where $\left| y \right|\ll O\left(1\right)$
the perturbations oscillate in the short time
and they emit the high energy photons~\cite{Horowitz:1978fq}
which is inconsistent with the cosmological observations.

Let us discuss how large gravitational curvature couplings are required. 
As we saw in the de Sitter or radiation-dominated Universe 
the smallness of $\left| y \right|$ amplifies the instabilities, and therefore, 
the higher-curvature couplings should satisfy,
\begin{align}\label{eq:stable-condition2}
\left| y\right| \gtrsim 1 \ \Longrightarrow \ 
{48\pi G_N }\left|\alpha_1\right|{ H }^{ 2 }_0\gtrsim 1 \, 
\ \Longrightarrow \ \left|\alpha_1\right|
\gtrsim 10^{118},
\end{align}
where we use the current Hubble parameter ${ H }_0$.
This requires large values of the gravitational curvature coupling 
$\left|a_{1,2,3}\right|\gtrsim 10^{118}$ or 
a large number of the particles species $\mathcal{N}\sim10^{118}$
for the theory.

%%%%%%%%%%%%%%%%%%%%%%%%%%%%%%%%%%%%%%%%%%%%%%%%
\section{Conclusion}
\label{sec:conclusion}
%%%%%%%%%%%%%%%%%%%%%%%%%%%%%%%%%%%%%%%%%%%%%%%%
In this work, we have discussed the spacetime instability 
in the effective field theories of quantum gravity, where the theory 
requires $R^{2}, { R }_{ \mu \nu  }{ R }^{ \mu \nu  }, R_{\mu\nu\kappa\lambda} 
R^{\mu\nu\kappa\lambda}$ as the leading quantum corrections. 
Although these higher-derivative curvatures are indispensable
for the renormalization of the semiclassical or quantum gravity theories, 
they lead to serious problems.
We have clearly shown 
that even if they are expressed as the Planck-suppressed operators 
they lead to catastrophic instability for the FLRW Universe.
The cosmological solutions for 
the effective field theories of gravity either grow exponentially 
or oscillate even in Planckian time $t_{\rm P}\approx 
\alpha_{ 1 }10^{-43}\ {\rm sec}$. 
In order to stabilize the universe 
the gravitational couplings must be rather large $\left|\alpha_{ 1 }\right|\gtrsim 10^{118}$ 
and a large number of the particles species $\mathcal{N}\sim10^{118}$ are expected 
at the current observed scale. 
Thus, the standard effective field theories of quantum gravity
naively fail to describe the observed Universe by the instability arising 
from higher-order derivative corrections. 

The best-known and only approach to solving this problem is to consider all higher-order derivative terms to be small perturbations. In the specific
perturbative approach~\cite{Simon:1990jn,Simon:1991bm,
Parker:1993dk,Anderson:2002fk,Frob:2013ht}, all higher-order differential terms, including local and nonlocal quantum corrections, running parameters, etc., are considered small perturbations to the Einstein-Hilbert terms of the GR, eliminating the instability problem due to ghosts. However, this is an ad hoc approach and it is not clear how valid it is. One would expect it to actually break down at larger energy scales, otherwise, it would prohibit the Starobinsky inflation model~\cite{Simon:1991bm}, which is phenomenologically very successful. Furthermore, the full contribution of higher-order derivative terms can not be ignored for sufficiently long timescales as our analysis. After all, the instability problem due to higher-order derivative terms in the effective field theories of quantum gravity is not exactly solved, and the directly derived modified Einstein equations contradict our observed universe.

%%%%%%%%%%%%%%%%%%%%%%%%%%%%%%%%%%%%%%%%%%%%%%%%
\section*{Acknowledgments}
%%%%%%%%%%%%%%%%%%%%%%%%%%%%%%%%%%%%%%%%%%%%%%%%
The author thanks Fuminobu Takahashi, Takahiro Terada and Naoki Watamura
for fruitful discussions and comments.

%%%%%%%%%%%%%%%%%%%%%%%%%%%%%%%%%%%%%%%%%%%%%%%%
\section*{Data Availability}
%%%%%%%%%%%%%%%%%%%%%%%%%%%%%%%%%%%%%%%%%%%%%%%%

The datasets generated and/or analyzed during the current study are available from the corresponding author on reasonable request.

%%%%%%%%%%%%%%%%%%%%%%%%%%%%%%%%%%%
%%%%%%%%%%%%%%%%%%%%%%%%%%%%
\appendix
\section{Geometrical tensors for FLRW spacetime}
%%%%%%%%%%%%%%%%%%%%%%%%%%%%%%%%%%%%%
%%%%%%%%%%%%%%%%%%%%%%%%%%
Here, we provide geometrical tensors for FLRW spacetime.
In this paper we take the FLRW line element as follows,
\begin{align}
ds^2=dt^2-a(t)^2\sum _{ i,j=1 }^{ 3 }{ { h }_{ ij }{ dx }^{ i }{ dx }^{ j } }, 
\end{align}
in which $a = a\left(t\right)$ express the scale factor with the cosmic time $t$ and,
\begin{align}
\sum _{ i,j=1 }^{ 3 }{ { h }_{ ij }{ dx }^{ i }{ dx }^{ j } }=
\frac { 1 }{ 1- {K}{ r }^{ 2 } } { dr }^{ 2 }+{ r }^{ 2 }\left( d{ \theta  }^{ 2 }+
\sin^{ 2 } { \theta d{ \phi  }^{ 2 } }  \right) , 
\end{align}
where $K $ is the spatial curvature parameter.
For simplicity, we consider spatially flat spacetime $K=0$.
The conformal time parameter $\eta$ is given by,
\begin{align}
d\eta=\frac { dt }{ a\left( t \right)  }
\end{align}
whose line element is given by
\begin{align}
ds^2=a^2(\eta)\left(d\eta^2-\sum _{ i,j=1 }^{ 3 }{ { h }_{ ij }{ dx }^{ i }{ dx }^{ j } }\right), 
\end{align}

We introduce $C(\eta)=a^2(\eta)$ and $D(\eta)=C(\eta)'/C(\eta)$
in which the prime $'$ express the derivative of $\eta$.
The Ricci tensor, Ricci scalar and other geometrical tensors
are given by~\cite{Birrell:1982ix},
\begin{align}
R_{00}&=\frac { 3 }{ 2 }D',\quad R_{11}=-\frac { 1 }{ 2 }\left(D'+D^{2}\right),
\quad R=\frac{3}{C}\left({ D' +\frac { 1 }{ 2 }  { D }^{ 2 } } \right),\\
H_{00}^{\left(1\right)}&=\frac { 9 }{ C }\left( -D''D+\frac { 1 }{ 2 }D'^{2} +\frac{3}{8}D^{4} \right), \\
H_{00}^{\left(3\right)}&=\frac { 3 }{ C }\left( \frac{1}{16}D^{4}\right).
\end{align}

%%%%%%%%%%%%%%%%%%%%%%%%%%%%%%%%%%%%%%%%%%%%%%%%
\section{Null energy condition and effective field theory of gravity}
\label{sec:null-energy}
%%%%%%%%%%%%%%%%%%%%%%%%%%%%%%%%%%%%%%%%%%%%%%%%
In  this appendix, we show that 
the higher-derivative gravitational corrections violate 
the null energy condition (NEC).
Although the NEC has little relation to the main result of this paper, 
we briefly mention it.
The NEC is the weakest but most standard energy conditions
to restrict the pathological spacetime for the 
general relativity and states that ${ T }_{ \mu\nu }$ satisfy, 
\begin{align}T_{\mu\nu}k^{\mu}k^{\nu}\ge  0,
\end{align}
for any null (light-like) vector $k^{\mu}$. 
For a perfect fluid with positive energy, 
the NEC yields the relation: $P+\rho \ge  0$.
The condition
preclude undesired consequences such as wormhole, spacetime instabilities, 
superluminal propagation and unitary violations~\cite{ 
Carroll:2003st,Cline:2003gs,Dubovsky:2005xd,Buniy:2006xf,Arefeva:2006ido} for general relativity and it is consistent with 
gravitational thermodynamics~\cite{ArkaniHamed:2007ky,Dubovsky:2006vk,
Eling:2007qd,Parikh:2015ret}.
It is widely believed that any physical system or theory should respect the condition
and the violation leads to pathology.
Now, we will discuss whether 
the effective field theory of quantum gravity with 
the action~(\ref{eq:gravity}) and (\ref{eq:higher}) violates the NEC
and the theoretical consequence.

For the cosmological framework of the flat FLRW universe,
the Friedmann equations yield a simple equation,
\begin{equation}\label{eq:Hubb}
\dot { H } =-4\pi { G }_{ N }\left( P+\rho  \right) \,.
\end{equation}
where the Hubble parameter decreases with time or stays constant
if the null energy condition $P+\rho \ge  0$ is satisfied.
Thus, the flat FLRW spacetime always decelerates
and finally terminates the expansion.
On the other hand,
the de Sitter spacetime is always stable 
and the cosmological constant $\Lambda$ 
satisfy the relation $P_{\Lambda}+\rho_{\Lambda} =  0$.
The time-evolution of the Universe can be classified as,
\begin{equation}
H\ \longrightarrow \ \begin{cases} 0 \quad\quad\quad  
\left( P+\rho >  0 \right) 
\\  {\rm const}\quad\ \left(P+\rho=  0 \right) \\  
\infty \quad\quad\ \ \left( P+\rho <  0 \right) \end{cases} \label{Hubble-d}
\end{equation}
For the slow-roll inflation driven by an inflaton field $\phi$, we have
$\dot{H} = -4\pi G_{N}\dot{\phi}^2 <0$
which is consistent with one's intuition.  
On the other hand, considering the 
ghost field which has negative kinetic terms and 
provides problems in QFT,
it leads to the relation $\dot{H} = 4\pi G_{N}\dot{\phi}^2 >0$.
Thus, we can expect that the Hubble expansion ratio
always decelerates or stays for ordinary cosmological theories.

First, let us consider semiclassical gravity which includes 
the backreaction effects of the quantum fluctuations onto spacetime.
The de Sitter spacetime can be interoperated as
one observer is surrounded by 
thermal radiation at the Hawking temperature $T_{\rm H}=H/2\pi$~\cite{Gibbons:1977mu} from the horizon.
The energy density or pressure including the thermal de Sitter radiation can be written as
\begin{equation}\label{eq:de Sitter radiation}
\rho_{\rm dS}=\rho_{\Lambda}+\frac{H^4}{480\pi^2}, \quad 
P_{\rm dS}=P_{\Lambda}+\frac{1}{3}\frac{H^4}{480\pi^2}\,.
\end{equation}
The backreaction of the thermal Hawking radiation satisfies the NEC:
$P_{\rm dS}+\rho_{\rm dS}\ge  0$
and 
terminates the expansion as follows~\cite{Markkanen:2017abw}
\begin{equation}\label{eqn:thermalHubble}
\dot { H }=-\frac { { G }_{ N }{ H }^{ 4 } }{ 720{ \pi  }^{ 2 } } <0
\ \Longrightarrow \ H=\frac { { H }_{ 0 }}{ { \left( \frac { { G }_{ N }{ H }_{0 }^{ 3 }t }
{ 240{ \pi  }^{ 2 } } +1 \right)  }^{ 1/3 } } \,.
\end{equation}
where it is not surprising that the thermal backreaction of 
Eq.~(\ref{eq:de Sitter radiation}) satisfies the NEC 
since we regard the quantum corrections as classic matters.
However, the above thermal interpretation of the de Sitter particle creations is not exact
and it is necessary for a detailed consideration based on the QFT approach.
In order to take into account of the gravitational vacuum polarization and 
quantum particle creation we usually consider the vacuum expectation values of
the energy-momentum tensor $\left< { T }_{ \mu \nu  } \right>$.
For a massless minimally coupled scalar field, 
the renormalized energy-momentum tensor is 
computed approximately for the Bunch-Davies vacuum as follows~\cite{Bunch:1978yw}:
\begin{align}
\begin{split}\label{eq:Bunch-Davies}
\left< { T }_{ \mu \nu  } \right>&= \frac{1}{2880\pi^2} 
\left(-\frac{1}{6}{ H }_{\mu \nu   }^{ \left( 1 \right)}
+ { H }_{ \mu \nu   }^{ \left( 3 \right)} \right) 
-\,\frac{{ H }_{ \mu \nu }^{ \left( 1 \right)}}{1152\pi^2}\log \left( \frac{R}{\mu^2} \right)  \\
&  +\frac{1}{13824\pi^2} \left[-32{ \nabla  }_{ \nu  }{ \nabla  }_{ \mu  }R
+56\Box Rg_{\mu\nu} - 8R R_{\mu\nu}+11R^2g_{\mu\nu} \right]
\end{split}
\end{align}
For simplicity, let us consider massless conformal coupled fields and 
the renormalized energy-momentum tensor for the scalar field 
is given by Eq.~(\ref{eq:conformal}).
The corresponding energy density or pressure is,
\begin{equation}
\rho_{\rm conformal}+p_{\rm conformal}=\frac{{ H }^{ 2 }\dot { H } }{720\pi^2}
+\frac{6\dot { H }^{ 2 } +3H\ddot { H } + \dddot { H }}{1440\pi^2} \gtrless  0 \,,
\end{equation}
which breaks the NEC and leads to the expansion $\dot{H} >0$~\cite{Matsui:2018iez}.
Although semiclassical gravity does not quantize the metric 
it takes into account
the backreaction of the quantum matter fields properly. 
However, the semiclassical gravity suffers from the spacetime instability 
and has the NEC violation.
Similarly, we show the relation of the NEC for 
the effective field theory of gravity with 
the action~(\ref{eq:gravity}) and (\ref{eq:higher}),
%%%%%%%%%%%%%%%%%%%%%%%%%%%%%
\footnote{
We note that the averaged null energy condition (ANEC)~\cite{Borde:1987qr}
is also violated.}
%%%%%%%%%%%%%%%%%%%%%%%%%%%%%
\begin{align}
\begin{split}\label{eq:dhjh-equation}
\dot { H }&={8\pi G_N }2\alpha_3{ H }^{ 2 }\dot { H }
-{8\pi G_N }{6\alpha_1}
\left(6\dot { H }^{ 2 } +3H\ddot { H } + \dddot { H }\right) \\
&-4\pi { G }_{ N }\left( P_{\rm matter}+\rho_{\rm matter}  \right) \gtrless  0\,,
\end{split}
\end{align}
where we can regard the higher-derivative gravitational corrections as quantum matter.
It is clear that 
the effective field theory violates the NEC from the higher-derivative 
gravitational corrections and the Hubble expansion ratio can increase.
We can similarly see unnatural consequences by using de Sitter entropy.
The character of the gravitational thermodynamics
in de Sitter spacetime is summarized by
the de Sitter entropy~\cite{Gibbons:1977mu} and the time-evolution is written as follows,
\begin{equation*}
\frac{dS_{\rm dS}}{dt} = -\frac{2\pi H^{-3}\dot{H}}{G_{N}}\ \Longleftrightarrow \ 
\frac{dS_{\rm dS}}{dN_{\rm tot}} = -\frac{2\pi H^{-4}\dot{H}}{G_{N}}
\end{equation*}
Considering the classical matter $P_{\rm matter}+\rho_{\rm matter}\ge  0$
the de Sitter entropy always increases,
\begin{equation*}
\frac{dS_{\rm dS}}{dt} = 8\pi^2 H^{-3}
\left(P_{\rm matter}+\rho_{\rm matter}\right) > 0.
\end{equation*}
On the other hand, for the effective field theory of gravity, the de Sitter entropy 
can decrease as follows,
\begin{align}\label{eqn:ddS}
\frac{dS_{\rm dS}}{dt} = 
-32\pi^2 \alpha_3{ H }^{ -1 }\dot { H }+
96\pi^2\alpha_1
\left(6{ H }^{-3}\dot { H }^{2 } +3H^{-2}\ddot { H } + H^{-3}\dddot { H }\right)\gtrless 0
\end{align}
which is inconsistent with gravitational thermodynamics~\cite{ArkaniHamed:2007ky}.

%%%%%%%%%%%%%%%%%%%%%%%%%%%%%%%%%%%%%%%%%%%%%%%%%%%%%%%%%%%%%

\bibliographystyle{JHEP}
\bibliography{reference}

\end{document}